\documentclass[aip,amsmath,amssymb,reprint,superscriptaddress,floatfix]{revtex4-1}

\usepackage[utf8]{inputenc}
\usepackage[T1]{fontenc}
\usepackage[english]{babel}
\usepackage{etoolbox}
\usepackage{graphicx}
\usepackage{xr}

\makeatletter
\newcommand*{\addFileDependency}[1]{%
  \typeout{(#1)}\@addtofilelist{#1}\IfFileExists{#1}{}{\typeout{No file #1.}}}
\makeatother

\newcommand*{\myexternaldocument}[2][]{%
  \externaldocument[#1]{#2}\addFileDependency{#2.tex}\addFileDependency{#2.aux}}
\myexternaldocument[S-]{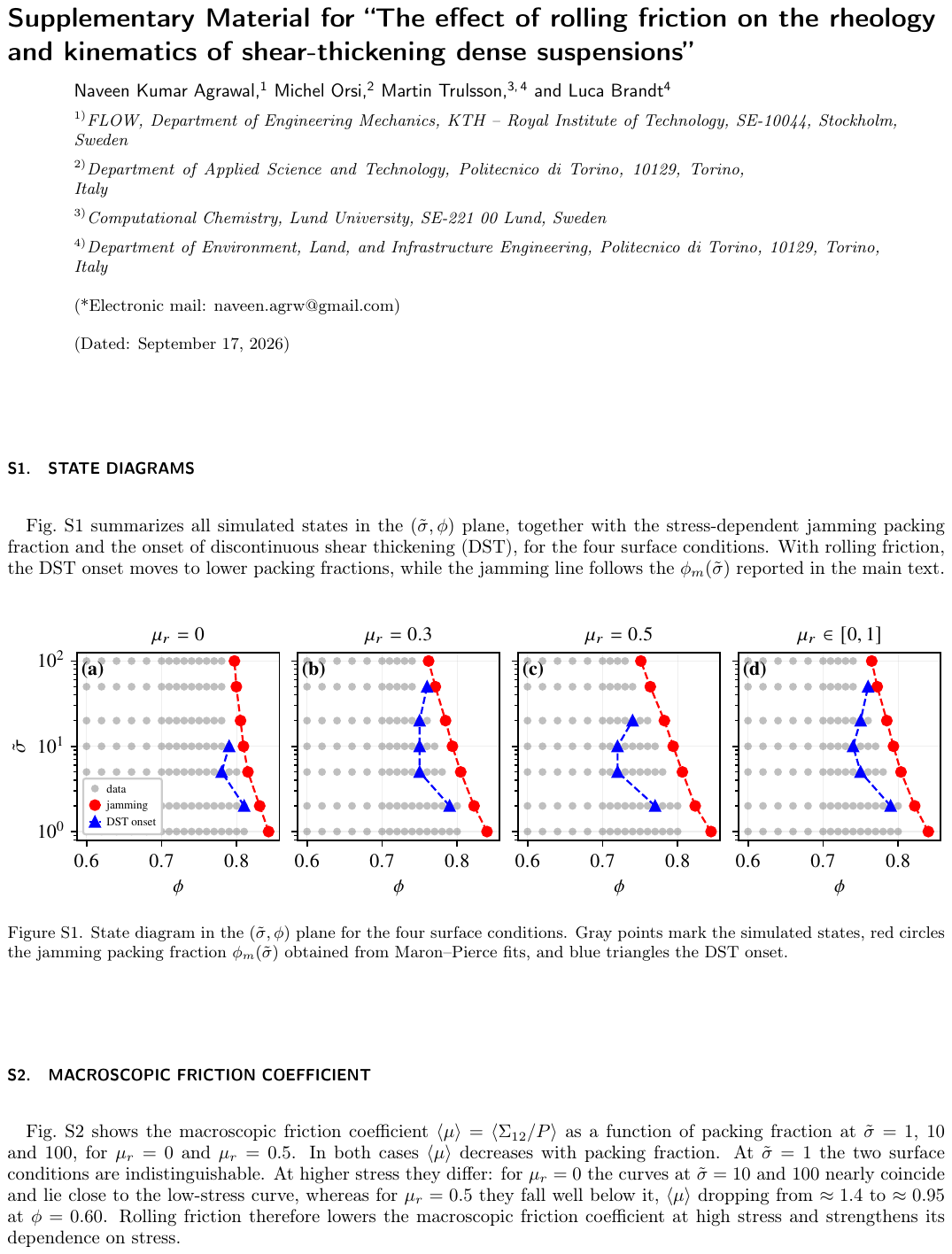}

\makeatletter
\def\@email#1#2{%
 \endgroup
 \patchcmd{\titleblock@produce}
  {\frontmatter@RRAPformat}
  {\frontmatter@RRAPformat{\produce@RRAP{*#1\href{mailto:#2}{#2}}}\frontmatter@RRAPformat}
  {}{}
}%
\makeatother

\begin{document}

\title{The effect of rolling friction on the rheology and kinematics of shear-thickening dense suspensions}

\author{Naveen Kumar Agrawal}
\email{naveen.agrw@gmail.com}
\affiliation{FLOW, Department of Engineering Mechanics, KTH -- Royal Institute of Technology, SE-10044, Stockholm, Sweden}

\author{Michel Orsi}
\affiliation{Department of Applied Science and Technology, Politecnico di Torino, 10129, Torino, Italy}

\author{Martin Trulsson}
\affiliation{Computational Chemistry, Lund University, SE-221 00 Lund, Sweden}
\affiliation{Department of Environment, Land, and Infrastructure Engineering, Politecnico di Torino, 10129, Torino, Italy}

\author{Luca Brandt}
\affiliation{Department of Environment, Land, and Infrastructure Engineering, Politecnico di Torino, 10129, Torino, Italy}

\date{\today}

\begin{abstract}
Sliding friction lowers the jamming packing fraction of dense suspensions and, when activated by stress, drives non-inertial shear thickening. Additional rolling resistance lowers the jamming packing fraction further and can represent effects of particle roughness or angularity. How it changes particle motion on the approach to jamming remains unclear. Using stress-controlled two-dimensional discrete-element simulations, we compare a sliding-only suspension with systems having either uniform or surface-varying rolling friction. At fixed packing fraction, adding rolling friction changes continuous shear thickening into discontinuous shear thickening. At the same distance from the stress-dependent jamming point, $\Delta\phi=\phi-\phi_m(\sigma)$, however, the flow curves nearly collapse, showing that the rheological effect arises largely from the shift in $\phi_m$. Guided by this collapse, we compare particle kinematics in the high-stress thickened state at matched $\Delta\phi$, revealing differences hidden by the similar bulk response. Translational velocity correlations extend over several particle diameters, whereas rotational correlations remain local. Rolling friction promotes co-rotation at contact in place of strong counter-rotation and suppresses rotational relative to translational fluctuations. Rotation nevertheless becomes increasingly important near jamming in every case. The suspension with surface-varying rolling friction follows the behavior of a uniform system with $\mu_r\approx0.3$ because the coefficients sampled at contacts lie well below the surface average value, here $\mu_r\approx0.5$. Thus, $\Delta\phi$ largely organizes the shear-thickening rheology, but not the particle kinematics. These retain a distinct signature of the rolling constraint that must be considered when rolling friction is used to model rough or angular particles.
\end{abstract}

\maketitle

\section{Introduction}\label{sec:introduction}

Dense suspensions consist of solid particles dispersed at high concentration in a fluid. They occur in industrial materials ranging from cement \citep{vandamme2018concrete} and pharmaceuticals \citep{stickel2005fluid} to food products \citep{blanco2019conching,orsi2026shear}, as well as in natural flows such as avalanches and landslides \citep{coussot1997mudflow}. Despite their simple composition, they exhibit yielding, shear thinning, shear thickening, and shear jamming \citep{cates1998jamming,brown2014shear,ness2021physics,morris2020shear,morris2023progress}. In the dense regime, these macroscopic responses are governed to a large extent by the constraints imposed at particle contacts.

Sliding friction provides the basic link between contact mechanics, shear thickening, and jamming. Frictionless particles jam at a higher packing fraction than particles whose relative tangential motion is constrained. If a short-range repulsion keeps particles apart at low stress but is overcome as the stress increases, the contact network changes from predominantly lubricated to frictional. The accompanying decrease in the stress-dependent jamming packing fraction produces non-inertial shear thickening and, at sufficiently high packing fraction, shear jamming \citep{seto2013discontinuous,mari2014shear,wyart2014discontinuous}. This mechanism underlies most current numerical descriptions of shear-thickening suspensions.

Quantitative discrepancies nevertheless remain between simulations based only on sliding friction and experiments. For frictionless spheres in three dimensions, the jamming packing fraction is approximately $\phi_m=0.64$, close to random close packing \citep{ohern2003jamming}. Increasing the sliding-friction coefficient $\mu_s$ lowers $\phi_m$, approaching approximately $0.56$ as $\mu_s\to\infty$ \citep{mari2014shear,singh2018constitutive}. In two dimensions, the corresponding values for bidisperse disks decrease from approximately $0.84$ in the frictionless limit to approximately $0.78$ at large $\mu_s$ \citep{ohern2003jamming,sharma2026frictional}. Although $\phi_m$ can therefore be adjusted by changing $\mu_s$, agreement with experiments may require sliding-friction coefficients larger than those measured independently. This suggests that sliding friction alone does not account for all the constraints present at the actual particle contacts.

Rolling friction supplies an additional constraint. When particles resist both sliding and rolling, the jamming packing fraction decreases further, strengthening shear thickening at fixed packing fraction and bringing simulations into better quantitative agreement with experiments \citep{singh2020shear,sharma2026frictional}. Rolling and twisting constraints have likewise been shown to reduce the packing fraction and coordination number of mechanically stable granular packings \citep{santos2020granular}. Although even more constraints can be introduced, their successive contributions become smaller, and sliding and rolling friction together already reproduce important features of suspension microstructure \citep{singh2022stress}.

Rolling friction is also useful as a reduced representation of particle shape. Industrial and natural particles are rarely perfectly smooth spheres; they are commonly rough, angular, faceted, elongated, or flattened. Such shapes alter packing, orientational order, rotational motion, and rheology \citep{estrada2011identification,trulsson2018rheology,bilotto2025shear,hsu2018roughness,hsu2021exploring,minten2025hydrodynamic}. Simulating them explicitly is considerably more expensive than simulating disks or spheres: particle orientations must be tracked, contacts must be detected between extended surfaces, and the hydrodynamic interaction between nearby non-spherical particles is more difficult to resolve. Applied to otherwise spherical particles, rolling resistance reproduces several mechanical effects associated with angularity, including a lower jamming packing fraction and a lower isostatic coordination number \citep{estrada2008shear,estrada2011identification,singh2020shear,santos2020granular}. It therefore offers a simple way to represent some consequences of roughness or shape without resolving the particle geometry itself.

Assigning one rolling-friction coefficient to an entire particle remains an idealization. A rough or faceted surface presents different local geometries, so its resistance to rolling depends on where a contact forms \citep{scherrer2024sliding}. The coefficient acting at a contact is then sampled from the two local surface values rather than prescribed as a single particle property. It is not evident whether the resulting suspension should behave according to the surface-averaged coefficient, the mean coefficient at contacts, or another feature of the contact-level distribution. We examine this question using a model in which $\mu_r$ varies periodically over each particle surface and comparing it with systems having uniform $\mu_r$. The model is deliberately minimal: it retains the hydrodynamic treatment of smooth particles while isolating the effect of spatially heterogeneous contact tribology.

The influence of rolling friction cannot be assessed from the jamming packing fraction alone. A disk in two dimensions has two translational and one rotational degree of freedom. The critical coordination number $Z_c$, defined as the mean number of contacts per particle required for isostaticity, is $Z_c=3$ when sliding is constrained but rolling remains free. Constraining rolling lowers it to $Z_c=2$ \citep{maxwell1864calculation,mari2019force,singh2020shear,santos2020granular,shundyak2007force}. Rotation is therefore part of the constraint counting that determines jamming, even though most diagnostics of shear jamming, including viscosity, translational velocity correlations, and network connectivity, emphasize translational motion \citep{seto2013discontinuous,bi2011jamming,santra2025rigid,pandare2026contact,goyal2024flow,vandernaald2024minimally}.

Recent work has shown that rotational statistics provide information not contained in these conventional measures. In suspensions with sliding friction alone, the non-affine angular-velocity distribution broadens as the contact network develops, rotations of nearby particles become increasingly anti-correlated, and rigid and non-rigid particles exhibit different rotational statistics \citep{orsi2026contact}. Because no constraint in that model acts directly on relative rotation, these signatures emerge from the organization of the contact network. The counter-rotation of neighboring particles is also the motion expected when two contacting particles roll past one another without rolling resistance.

Introducing rolling friction changes the contact law directly by opposing relative rotation. This raises two distinct questions. The first is rheological: does rolling resistance change shear thickening in a way that cannot be reduced to its shift of $\phi_m$? The second is kinematic: at a fixed distance from jamming, how does the rolling constraint redistribute non-affine motion between translation and rotation, and how does it alter the spatial organization of those motions?

We address these questions using stress-controlled simulations with a fixed sliding-friction coefficient $\mu_s=1$. We compare a sliding-only reference, $\mu_r=0$, with uniform rolling-friction coefficients $\mu_r=0.3$ and $0.5$, and with a system in which $\mu_r$ varies over each particle surface between $0$ and $1$. We first determine how rolling friction changes the flow curves and the stress-dependent jamming packing fraction. We then compare the systems at matched distance from jamming and examine the frictional coordination number, angular-velocity distributions, translational and rotational velocity correlations, and translational and rotational granular temperatures. This separation between rheology and kinematics allows us to determine whether states with similar bulk flow responses also have similar particle dynamics.

The remainder of the paper is organized as follows. Section~\ref{sec:model} describes the numerical method, contact laws, surface-varying rolling-friction model, and simulation protocol. Section~\ref{sec:results} presents the bulk rheology, contact statistics, and particle kinematics. Section~\ref{sec:conclusion} summarizes the main findings.

\section{Numerical Model}\label{sec:model}

We simulate non-Brownian suspensions in the Stokes regime, with particle Reynolds number $Re_p=0$ and Péclet number $Pe=\infty$. Particle and fluid inertia are therefore neglected, and each particle satisfies instantaneous force and torque balance. The code builds on a previously published solver \citep{orsi2022simulation,orsi2023frame,orsi2025effect}. Here we use its discrete-element-method (DEM) branch, in which the hydrodynamic interactions are represented by pairwise near-field lubrication and single-particle Stokes drag, as commonly done in reduced simulations of dense suspensions \citep{seto2013discontinuous,mari2014shear}. Contact and short-range repulsive forces complete the model.

For each particle $i$, the force and torque balances are
\begin{subequations}
\begin{equation}
    \mathbf 0
    =
    \mathbf F_i^{\mathrm H}
    + \mathbf F_i^{n}
    + \mathbf F_i^{t}
    + \mathbf F_i^{\mathrm{rep}} \ ,
\end{equation}
\begin{equation}
    \mathbf 0
    =
    \mathbf T_i^{\mathrm H}
    + \mathbf T_i^{s}
    + \mathbf T_i^{r} \ .
\end{equation}
\end{subequations}
Here, $\mathbf F_i^{\mathrm H}=\mathbf F_i^{\mathrm{lub}}+\mathbf F_i^{\mathrm{drag}}$ and $\mathbf T_i^{\mathrm H}=\mathbf T_i^{\mathrm{lub}}+\mathbf T_i^{\mathrm{drag}}$ collect the lubrication and drag contributions. The terms $\mathbf F_i^{n}$, $\mathbf F_i^{t}$, and $\mathbf F_i^{\mathrm{rep}}$ are the normal-contact, tangential-contact, and repulsive forces, while $\mathbf T_i^{s}$ and $\mathbf T_i^{r}$ are the torques generated by sliding and rolling friction. Each contact contribution vanishes when the corresponding interaction is inactive. The lubrication force and torque are computed from the standard pairwise resistance functions for two unequal spheres \citep{jeffrey1984calculation,jeffrey1992calculation}. The drag force and torque follow the single-particle Stokes laws relative to the local undisturbed flow, $\mathbf u^\infty(\mathbf R_i)$ and $\boldsymbol\omega^\infty(\mathbf R_i)$.

For particles $i$ and $j$, with radii $a_i$ and $a_j$ and positions $\mathbf R_i$ and $\mathbf R_j$, we define $r_{ij}=\lvert\mathbf R_j-\mathbf R_i\rvert$, $h_{ij}=r_{ij}-(a_i+a_j)$, and $\mathbf n_{ij}=(\mathbf R_j-\mathbf R_i)/r_{ij}$. The vector $\mathbf n_{ij}$ points from $i$ to $j$. We write $\mathbf F_{ij}$ for the force exerted by $j$ on $i$, so that $\mathbf F_{ji}=-\mathbf F_{ij}$. Contact is activated when $h_{ij}<h_r$, where $h_r=0.005\,a_s$ and $a_s$ is the radius of the smaller particle species. The overlap relative to this contact surface is $\delta_n=h_r-h_{ij}$, and the normal force is Hertzian and repulsive:
\begin{equation}
    \mathbf F_{ij}^{n}
    =
    -k_n\,\delta_n^{3/2}\mathbf n_{ij} \ ,
    \quad k_n>0 \ .
    \label{eq:normal}
\end{equation}
Because contact occurs at the positive gap $h_r$, the lubrication resistance remains finite when the contact force is activated; no additional lubrication regularization is required at contact. Particle size and packing fraction are defined using the geometric radii $a_i$, without adding $h_r$ to the particle dimensions. Including this small roughness length in the effective radii would shift the reported packing fractions by an amount of order $h_r/a_s$, without changing the physical trends. The stiffness $k_n$ is chosen sufficiently large that particle overlaps remain negligible on the scale of $a_s$.

Sliding friction is represented by an elastic tangential spring capped by a Coulomb condition. Its displacement $\boldsymbol\xi_{ij}^{t}$ evolves according to the relative tangential velocity of the two surfaces at contact:
\begin{equation}
    \dot{\boldsymbol\xi}_{ij}^{t}
    =
    \left(\mathbf U_i-\mathbf U_j\right)_{\!\perp}
    +
    \left(
        a_i\boldsymbol\omega_i
        +
        a_j\boldsymbol\omega_j
    \right)\times\mathbf n_{ij} \ ,
    \label{eq:slip-t}
\end{equation}
where $\mathbf v_\perp =\mathbf v-(\mathbf v\cdot\mathbf n_{ij})\mathbf n_{ij}$ denotes projection onto the contact tangent. The tangential force is
\begin{equation}
    \mathbf F_{ij}^{t}
    =
    \begin{cases}
        -k_t\,\boldsymbol\xi_{ij}^{t} \ ,
        &
        k_t\lvert\boldsymbol\xi_{ij}^{t}\rvert
        \leq
        \mu_s\lvert\mathbf F_{ij}^{n}\rvert \ ,
        \\
        -\mu_s\lvert\mathbf F_{ij}^{n}\rvert
        \dfrac{\boldsymbol\xi_{ij}^{t}}
        {\lvert\boldsymbol\xi_{ij}^{t}\rvert} \ ,
        &
        \text{otherwise} \ ,
    \end{cases}
    \label{eq:sliding}
\end{equation}
where $\mu_s$ is the sliding-friction coefficient. The tangential stiffness is $k_t=\frac{2}{7}\frac{\lvert\mathbf F_{ij}^{n}\rvert}{\delta_n}$, as commonly used in DEM models of granular contacts. The corresponding torques are
\begin{equation}
    \mathbf T_{ij}^{s}
    =
    a_i\,\mathbf n_{ij}\times\mathbf F_{ij}^{t} \ ,
    \quad
    \mathbf T_{ji}^{s}
    =
    a_j\,\mathbf n_{ij}\times\mathbf F_{ij}^{t} \ .
    \label{eq:sliding-torque}
\end{equation}

Rolling friction opposes relative rotation of the contacting particles \citep{dominik1995resistance,krijt2014rolling}. Following \citet{luding2008cohesive,singh2020shear,sharma2026frictional}, we use the same spring-and-Coulomb construction as for sliding friction, but with the reduced radius $\widetilde a_{ij}=(a_i a_j)/(a_i+a_j)$. The rolling-spring displacement evolves as
\begin{equation}
    \dot{\boldsymbol\xi}_{ij}^{r}
    =
    \widetilde a_{ij}
    \left(
        \boldsymbol\omega_i-\boldsymbol\omega_j
    \right)\times\mathbf n_{ij} \ ,
    \label{eq:slip-r}
\end{equation}
and the rolling quasi-force is
\begin{equation}
    \mathbf F_{ij}^{r}
    =
    \begin{cases}
        -k_t\,\boldsymbol\xi_{ij}^{r} \ ,
        &
        k_t\lvert\boldsymbol\xi_{ij}^{r}\rvert
        \leq
        \mu_r\lvert\mathbf F_{ij}^{n}\rvert \ ,
        \\
        -\mu_r\lvert\mathbf F_{ij}^{n}\rvert
        \dfrac{\boldsymbol\xi_{ij}^{r}}
        {\lvert\boldsymbol\xi_{ij}^{r}\rvert} \ ,
        &
        \text{otherwise} \ .
    \end{cases}
    \label{eq:rolling}
\end{equation}
Here $\mu_r$ is the rolling-friction coefficient. The quasi-force $\mathbf F_{ij}^{r}$ generates no net translational force and enters only through the torque
\begin{equation}
    \mathbf T_{ij}^{r}
    =
    \widetilde a_{ij}
    \mathbf n_{ij}\times\mathbf F_{ij}^{r}
    =
    -\mathbf T_{ji}^{r} \ .
    \label{eq:rolling-torque}
\end{equation}
Every contact is subject to both sliding and rolling friction from the moment it forms. This differs from critical-load models, in which friction itself is activated only when the normal load exceeds a prescribed threshold \citep{seto2013discontinuous,mari2014shear}.

A short-range repulsive force introduces the stress scale required for shear thickening:
\begin{equation}
    \mathbf F_{ij}^{\mathrm{rep}}
    =
    -F_0
    \exp\!\left(-h_{ij}/\lambda\right)
    \mathbf n_{ij} \ ,
    \label{eq:repulsion}
\end{equation}
where $F_0$ is the repulsive force scale and $\lambda=0.05\,a_s$ is the interaction range. The absolute value of $F_0$ only fixes the stress unit. In two dimensions, the corresponding stress scale is $\sigma_0=F_0/a_s$ and the imposed stress is reported as $\tilde\sigma=\sigma/\sigma_0$. Both contact (Eq.~\eqref{eq:normal}) and repulsive (Eq.~\eqref{eq:repulsion}) forces are capped for $h_{ij} \le 10^{-5}$; both lubrication and repulsive forces are truncated for $h_{ij} > 0.3a_s$. At low $\tilde\sigma$, the repulsion prevents sustained frictional contact. Increasing $\tilde\sigma$ allows the imposed forces to overcome the repulsive barrier, activating frictional contacts and producing shear thickening.

\subsection{Variable rolling friction}\label{sec:varmur}

A rough or faceted particle need not offer the same resistance to rolling at every point on its surface \citep{ruiz-lopez2023tribological}. To represent this heterogeneity without resolving the surface geometry, we prescribe a rolling-friction coefficient that varies with the angular position $\theta_i\in[0,2\pi)$ on particle $i$:
\begin{equation}
    \mu_r^{(i)}
    =
    \frac{1}{2}
    \left[
        \sin(4\theta_i)+1
    \right] \ .
    \label{eq:varmur}
\end{equation}
This fourfold texture spans the interval $\mu_r^{(i)}\in[0,1]$ and has a surface average of $0.5$.

The probability of finding a local coefficient $m\equiv\mu_r^{(i)}$ at a uniformly sampled surface position follows the arcsine distribution on $[0,1]$:
\begin{equation}
    P(m)
    =
    \frac{1}{\pi\sqrt{m(1-m)}} \ .
    \label{eq:arcsine}
\end{equation}
Thus, the prescribed surface contains more points with coefficients near $0$ or $1$ than near its mean value $0.5$.

A contact samples one coefficient from each of the two participating surfaces. We define the rolling-friction coefficient acting between particles $i$ and $j$ as
\begin{equation}
    \mu_r^{ij}
    =
    \min\!\left(
        \mu_r^{(i)},
        \mu_r^{(j)}
    \right) \ ,
    \label{eq:minrule}
\end{equation}
so that the smoother of the two local surface regions limits the resistance to rolling. This minimum rule is a modeling choice rather than a unique consequence of contact mechanics. Its advantage is that the corresponding contact-level distribution can be derived analytically.

If contact locations are sampled independently and uniformly from the two surfaces, which we call random pairing, then $\mu_r^{(i)}$ and $\mu_r^{(j)}$ are independent draws from Eq.~\eqref{eq:arcsine}. The probability density function (PDF) of their minimum is
\begin{equation}
    P_{\mathrm{contact}}(m)
    =
    2\left[1-F(m)\right]P(m) \ ,
    \label{eq:pcontact}
\end{equation}
where $F(m)=\frac{2}{\pi}\arcsin\!\left(\sqrt{m}\right)$ is the cumulative distribution associated with $P(m)$. The mean of the contact-level distribution is
\begin{equation}
    \left\langle\mu_r^{ij}\right\rangle_{\mathrm{contact}}
    =
    \frac{1}{2}-\frac{2}{\pi^2}
    \approx0.297 \ .
    \label{eq:murmean}
\end{equation}
The minimum rule therefore gives a mean contact coefficient well below the surface average $0.5$.

Other combination rules would select larger contact coefficients. Under random pairing, the geometric mean has expectation $4/\pi^2\approx0.405$, while the arithmetic mean has expectation $0.5$. Since $\min(a,b)\leq\sqrt{ab}\leq(a+b)/2$, neither rule can produce a lower coefficient than the minimum for a given pair. We do not simulate these alternative rules. If the response is governed primarily by the coefficients realized at contacts, as the results in Sec.~\ref{sec:mur_eff} indicate, they should produce behavior corresponding to a larger effective $\mu_r$.

The random-pairing result also lies between two limiting forms of contact selection. If contacts always joined surface points with equal coefficients, $\mu_r^{(i)}=\mu_r^{(j)}$, the minimum rule would return the surface average $0.5$. If they always joined complementary values, $\mu_r^{(j)}=1-\mu_r^{(i)}$, the mean would be $1/2-1/\pi\approx0.182$. Equation~\eqref{eq:murmean} therefore provides a parameter-free benchmark against which the measured contact distribution can be compared. We have also verified that applying the same type of surface variation to the sliding-friction coefficient produces analogous selection by the coefficients realized at contacts, rather than by their surface average.

\subsection{Numerical setup}\label{sec:simulation}

The simulations are two-dimensional and use Lees--Edwards periodic boundary conditions under simple shear. Two-dimensional models reproduce the main shear-thickening and contact-network phenomenology of three-dimensional dense suspensions when the packing fraction is measured relative to the appropriate jamming point \citep{santra2025rigid,pandare2026contact,sharma2026frictional}. In the present geometry, $\phi$ denotes the particle area fraction.

The simulations are stress-controlled. We impose $\tilde\sigma=\sigma/\sigma_0$, with $\sigma_0=F_0/a_s$, and determine the instantaneous shear rate from the force and stress balances. Each simulation is continued to a total accumulated strain $\gamma=12$. Steady state is normally reached within approximately two units of strain, after which the rheological and particle-scale quantities are sampled. The relative viscosity is $\eta_r=\frac{\sigma}{\eta_0\dot\gamma}$, where $\eta_0$ is the viscosity of the suspending fluid.

We use bidisperse disks with radii $a_s=1$ and $a_l=1.4$ to suppress crystallization. These geometric radii, excluding the roughness length $h_r$, are used to calculate $\phi$. The sliding-friction coefficient is fixed at $\mu_s=1$ throughout. The four surface conditions are therefore distinguished only by their rolling friction: $\mu_r=0$, $\mu_r=0.3$, $\mu_r=0.5$, and $\mu_r^{(i)}\in[0,1]$. The first case is the sliding-only reference. The second and third have uniform rolling-friction coefficients, while the last uses the surface variation defined in Eq.~\eqref{eq:varmur}. All remaining hydrodynamic, contact, and repulsive-force parameters are held fixed.

Each surface condition is simulated over packing fractions $\phi\in[0.60,0.82]$ and imposed stresses $\tilde\sigma=[1,2,5,10,20,50,100]$.

\begin{figure*}[t]
    \centering
    \includegraphics[width=\linewidth]{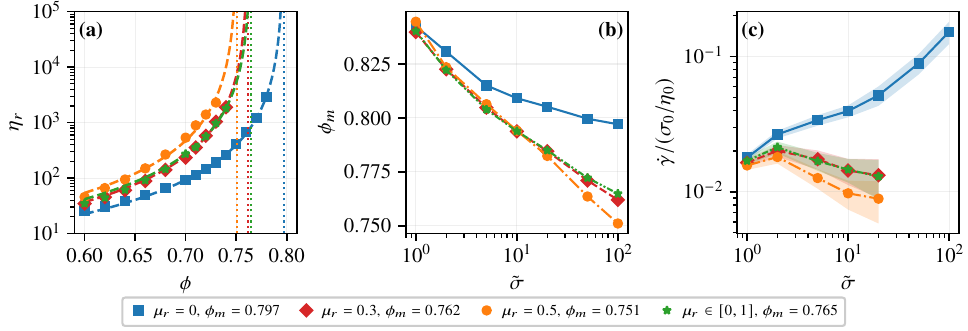}
    \caption{Rheology of suspensions with uniform and surface-varying rolling resistance, $\mu_r=0$, $0.3$, $0.5$, and $\mu_r\in[0,1]$ variable across each particle surface. (a) Relative viscosity $\eta_r$ versus packing fraction $\phi$ at $\tilde\sigma=100$; dashed lines show Maron--Pierce fits and dotted vertical lines the extrapolated jamming packing fractions $\phi_m$ listed in the legend. (b) Extrapolated jamming packing fraction $\phi_m$ versus imposed stress $\tilde\sigma$. (c) Shear rate versus imposed stress at $\phi=0.76$; shaded bands show the standard deviation of the shear rate in steady state.}
    \label{f:rheology}
\end{figure*}

\begin{figure*}[t]
    \centering
    \includegraphics[width=\linewidth]{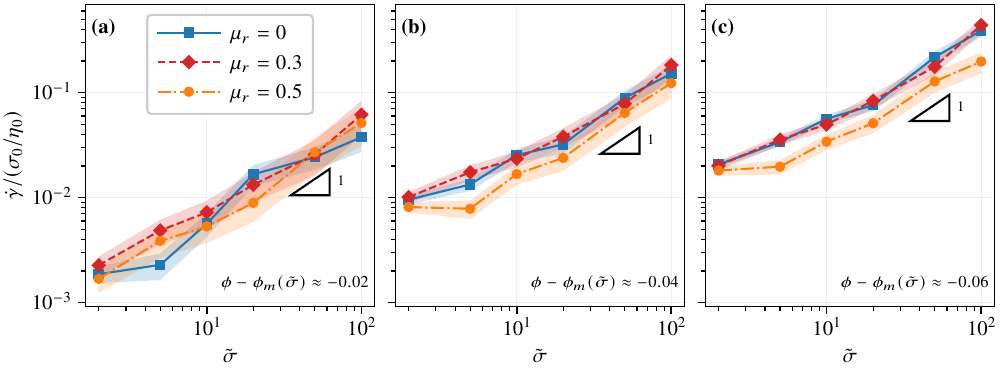}
    \caption{Shear rate versus imposed stress at three fixed distances from jamming, $\phi-\phi_m(\tilde\sigma)\approx$ (a) $-0.02$, (b) $-0.04$, and (c) $-0.06$, for $\mu_r=0$, $0.3$, and $0.5$; each point corresponds to a different packing fraction. Shaded bands show the standard deviation of the shear rate in steady state. Triangles indicate a slope of unity.}
    \label{f:flow_curve_fixed_gap}
\end{figure*}

\section{Results}\label{sec:results}

We first characterize the bulk rheology, establishing how rolling resistance shifts the jamming packing fraction and alters the character of shear thickening. We then descend to the particle level starting by measuring the frictional coordination number. From there, we turn to the kinematics while approaching shear jamming, examining in turn the distribution of the non-affine angular velocity, the spatial correlations of the non-affine translational and rotational velocities, and the translational and rotational granular temperatures. Finally, we determine the effective rolling-friction coefficient realized at a contact in the variable-$\mu_r$ suspension from the contact statistics, and show that it accounts for the rheological and kinematic behavior reported above.

\subsection{Bulk rheology}

Fig.~\ref{f:rheology} summarizes the rheological response for the three uniform rolling-friction cases and for the variable-$\mu_r$ suspension, in which the coefficient varies across each particle surface.

Fig.~\ref{f:rheology}(a) shows the relative viscosity as a function of packing fraction at $\tilde{\sigma}=100$, diverging near the jamming packing fraction $\phi_m$ following the Maron--Pierce correlation law $\eta_r = \alpha(\tilde{\sigma})\left[1-\phi/\phi_m(\tilde{\sigma})\right]^{-2}$, widely used to describe frictional suspension rheology \citep{maron1956application,morris2020shear,guy2015towards}. Rolling resistance shifts the divergence to lower packing fraction, from $\phi_m\approx0.80$ at $\mu_r=0$ to $\approx0.76$ at $\mu_r=0.3$ and $\approx0.75$ at $\mu_r=0.5$, consistent with the reduction reported for rolling-frictional suspensions \citep{singh2020shear,sharma2026frictional}. Fitting the same law at every imposed stress gives the stress-dependent jamming packing fraction shown in Fig.~\ref{f:rheology}(b). At the lowest stress, $\tilde\sigma=1$, all four surface conditions share $\phi_m\approx0.84$, close to the frictionless jamming point of bidisperse disks \citep{ohern2003jamming}. As the stress increases, $\phi_m$ decreases in every case, but more steeply with rolling friction: the rolling-frictional cases separate from the sliding-only case already at $\tilde\sigma=2$ and remain close to one another up to $\tilde\sigma\approx20$, beyond which the $\mu_r=0.5$ case falls lowest. The variable-$\mu_r$ suspension follows the $\mu_r=0.3$ case throughout. Notably, $\phi_m$ for the sliding-only suspension begins to plateau at lower stress than for the rolling-frictional cases, which have not yet saturated even at $\tilde\sigma=100$.

Fig.~\ref{f:rheology}(c) shows how rolling resistance changes the flow curve at fixed packing fraction $\phi=0.76$. Without rolling resistance the suspension undergoes continuous shear thickening (CST) at this packing fraction, whereas with rolling resistance it undergoes discontinuous shear thickening (DST), as evidenced by the negative slopes in the flow curves beyond $\tilde{\sigma}\approx2$, consistent with the promotion of DST by rolling friction reported by \citet{singh2020shear}. The shaded bands, which show the standard deviation of the steady-state shear rate, remain narrow across the entire stress range for $\mu_r=0$, but broaden markedly with increasing stress once rolling resistance is present. Such large fluctuations are characteristic of DST \citep{goyal2024flow,sedes2020fluctuations}.

The rolling-friction curves terminate at $\tilde{\sigma}=20$: at $\tilde{\sigma}=50$ the suspension is shear jammed. At this packing fraction, rolling resistance therefore lowers the stress required to reach shear jamming. A fixed packing fraction does not correspond to a fixed distance from jamming: $\phi_m$ decreases with increasing stress (Fig.~\ref{f:rheology}(b)), and rolling resistance lowers it further at all but the lowest stress, so that the rolling-frictional cases sit closer to jamming at $\phi=0.76$ than the sliding-only case. The character of shear thickening is known to depend on this distance from jamming \citep{wyart2014discontinuous,mari2015nonmonotonic,singh2018constitutive,ramaswamy2023universal}. We therefore compare the surface conditions at the same distance from jamming in Fig.~\ref{f:flow_curve_fixed_gap}, at $\phi-\phi_m(\tilde\sigma)\approx-0.02$, $-0.04$ and $-0.06$. Because $\phi_m$ depends on the stress, each point along these curves corresponds to a different packing fraction. At matched distance from jamming, the three uniform cases now lie close together, in contrast with their clear separation at fixed $\phi$, suggesting that the effect of rolling resistance on the flow curve at fixed $\phi$ arises largely through its shift of $\phi_m$. The simulated states are summarized in the $(\tilde\sigma, \phi)$ plane in supplementary material, Fig.~S1, together with the jamming line $\phi_m(\tilde\sigma)$ and the onset of DST.

The variable-$\mu_r$ suspension behaves like a suspension of uniformly rough particles with $\mu_r\approx0.3$ as outlined in Sec.~\ref{sec:varmur}: its viscosity curve is indistinguishable from the $\mu_r=0.3$ case across the full range of packing fractions, its jamming packing fraction follows the $\mu_r=0.3$ case at every stress, and its flow curve at $\phi=0.76$ follows the same branch. It does not behave like the suspension with uniform $\mu_r=0.5$, although the rolling-friction coefficient averaged over each particle surface is exactly $0.5$; the origin of this is discussed in Sec.~\ref{sec:mur_eff}.

The macroscopic friction coefficient $\langle\mu\rangle=\langle\Sigma_{12}/P\rangle$ is reported in the supplementary material, Fig.~S2. It decreases with packing fraction in all cases; at the lowest stress it is insensitive to rolling friction, whereas at higher stress rolling friction lowers it and strengthens its dependence on stress.

The results above show the effect of rolling friction on the bulk rheology. We now descend to the particle level, beginning with the contact network on approach to jamming.

\subsection{Frictional coordination number}\label{subsec:znet}

\begin{figure}[!t]
    \centering
    \includegraphics[width=\linewidth]{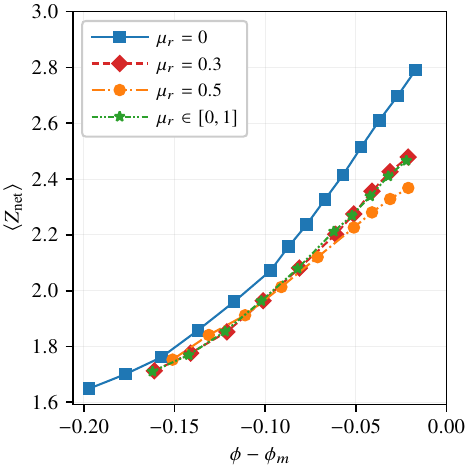}
    \caption{Frictional coordination number $Z_{\rm net}$, the mean number of frictional contacts per particle, versus distance from jamming $\phi-\phi_m$ at $\tilde\sigma=100$.}
    \label{f:znet}
\end{figure}

Since distance from jamming appears to be the relevant quantity regardless of stress, we focus here on $\tilde\sigma=100$, where differences are greatest. Fig.~\ref{f:znet} shows the frictional coordination number $Z_{\rm net}$, the mean number of frictional contacts per particle considering only particles with at least one contact, against the distance from jamming at $\tilde\sigma=100$. In all four cases $Z_{\rm net}$ grows as jamming is approached. Far from jamming the four surface conditions lie close together, but nearer $\phi_m$ they separate: the sliding-only suspension reaches the highest coordination, $\approx2.8$ at the closest approach we resolve, against $\approx2.5$ at $\mu_r=0.3$ and $\approx2.4$ at $\mu_r=0.5$. Rolling resistance therefore allows the suspension to approach jamming with fewer frictional contacts, as expected for an additional constraint at each contact that naturally lowers the jamming packing fraction \citep{estrada2011identification,sharma2026frictional}. The variable-$\mu_r$ suspension again follows the $\mu_r=0.3$ case. We now turn from the contact network to the particle-level kinematics, and to how rolling friction shapes them on approach to jamming.

\subsection{Angular velocity distribution}

We characterize the rotational motion through the non-affine angular velocity, obtained by removing the affine background, $\tilde{\omega} = \omega - \tfrac{1}{2}\dot{\gamma}$, where $\dot{\gamma}$ is the instantaneous shear rate. In two dimensions the angular velocity has only an out-of-plane component, and we drop the $z$ subscript throughout for brevity. To measure the fluctuation relative to the imposed flow, we use the normalized form $\omega' = \tilde{\omega}/\dot{\gamma}$. Throughout, tildes denote non-affine velocities and primes their normalized counterparts, the same convention applying to the translational components $v_x'=\tilde{v}_x/(\dot{\gamma}a_s)$ and $v_y'=\tilde{v}_y/(\dot{\gamma}a_s)$. For all four surface conditions, the mean rotation follows the affine value, $\langle\omega\rangle\approx\tfrac{1}{2}\dot{\gamma}$, so that $\langle\omega'\rangle\approx0$ and the distributions below describe fluctuations about the affine rotation. The distribution of $\omega'$ measures how strong these rotational fluctuations are, complementing the spatial correlations examined below.

\begin{figure}[!t]
    \centering
    \includegraphics[width=\linewidth]{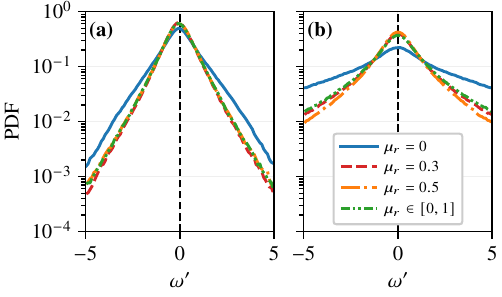}
    \caption{Probability density function of the normalized non-affine angular velocity $\omega'$ at $\tilde\sigma=100$ for the four surface conditions, (a) far from jamming, $\phi-\phi_m\approx-0.12$, and (b) close to jamming, $\phi-\phi_m\approx-0.025$. Rolling resistance narrows the distribution and suppresses the tails.}
    \label{f:wz_pdf_gap}
\end{figure}

Fig.~\ref{f:wz_pdf_gap} shows the PDF of the normalized angular velocity, $P(\omega')$, at $\tilde\sigma=100$ for the four surface conditions, at two distances from jamming, $\phi-\phi_m\approx-0.12$ (a) and $\approx-0.025$ (b). At both distances the sliding-only suspension has the broadest distribution, with the lowest central peak and the heaviest tails. Rolling friction narrows the distribution and suppresses the tails: far from jamming the three rolling-frictional cases nearly coincide, differing only in the far tails, while close to jamming the distribution narrows further with increasing $\mu_r$. The variable-$\mu_r$ suspension again follows the $\mu_r=0.3$ case. Rolling resistance therefore suppresses non-affine rotational motion, as expected of a constraint on the relative rotation of particles in contact.

\begin{figure*}[!t]
    \centering
    \includegraphics[width=\linewidth]{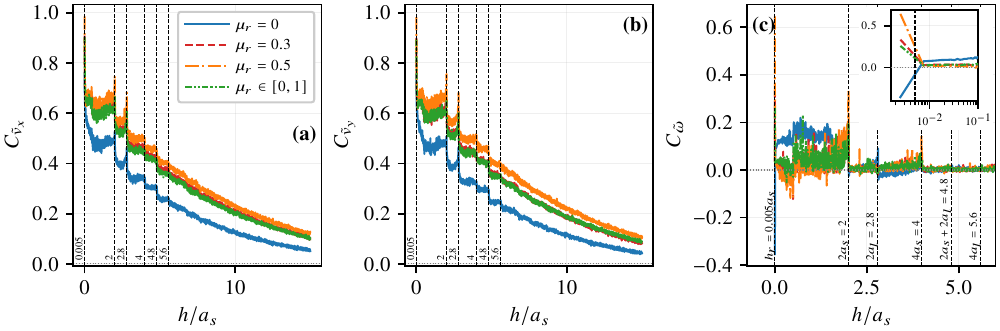}
    \caption{Spatial correlation of the non-affine velocities at $\tilde\sigma=100$. Panels show (a) the flow component $C_{\tilde{v}_x}$, (b) the flow-gradient component $C_{\tilde{v}_y}$ and (c) the rotational component $C_{\tilde{\omega}}$. For rolling friction coefficients $\mu_r=0$, $0.3$, $0.5$ and the variable case $\mu_r\in[0,1]$; close to jamming, $\phi-\phi_m \approx -0.025$. The dashed vertical lines correspond to gaps: $h_r=0.005a_s$, $2a_s=2$, $2a_l=2.8$, $4a_s=4$, $2a_s+2a_l=4.8$ and $4a_l=5.6$.}
    \label{f:c_v_w_h}
\end{figure*}

Comparing the two panels shows how the distribution evolves as jamming is approached. Measured relative to the shear rate, it broadens in all four cases, the central peak falling and the tails filling in. The tails rise by more than an order of magnitude at the edge of the plotted range, $|\omega'|=5$, in every case, so that an increasing fraction of particles rotate at rates far from the affine value as jamming is approached, as also found in sliding-frictional suspensions \citep{orsi2026contact}. The loss of the central peak, however, depends on the rolling constraint: between $\phi-\phi_m\approx-0.12$ and $-0.025$ it falls by about a factor of two for $\mu_r=0$, against about a factor of $1.5$ with rolling friction. Normalized rotational fluctuations therefore grow fastest when the relative rotation at contact is unconstrained.

\subsection{Velocity correlations}

The particle collective motion is characterized by the two-point spatial correlations of the non-affine velocities. The affine background is removed $\tilde{v}_x = v_x - \dot{\gamma}\,r_y$ and $\tilde{v}_y = v_y$, alongside the non-affine angular velocity $\tilde{\omega}$ introduced above. Pairs are binned by their surface gap $h_{ij}=r_{ij}-(a_i+a_j)$ rather than by their center-to-center distance, so that particle pairs of any size combination with the same gap fall in the same bin; contact corresponds to $h<h_r$, with $h_r=0.005\,a_s$ the roughness length introduced above. For each component we compute the equal-time, variance-normalized correlation 
\begin{equation}
    C_{\tilde{v}_x}(h) = \frac{\bigl\langle \tilde{v}_{x,i}\,\tilde{v}_{x,j}\bigr\rangle_h} {\bigl\langle \tilde{v}_x^{2}\bigr\rangle} \ ,
    \quad
    C_{\tilde{\omega}}(h) = \frac{\bigl\langle \tilde{\omega}_{i}\,\tilde{\omega}_{j}\bigr\rangle_h} {\bigl\langle \tilde{\omega}^{2}\bigr\rangle} \ ,
\end{equation}
where $\langle\cdot\rangle_h$ denotes an average over all pairs whose gap falls in the bin centered at $h$ ($C_{\tilde{v}_y}(h)$ is defined analogously). Normalizing by the variance removes the overall magnitude of the fluctuations, which varies strongly with $\phi$ and $\tilde{\sigma}$, and isolates the suspension spatial structure.

The translational correlations are examined first; see panels (a) and (b) of Fig.~\ref{f:c_v_w_h}. For all four surface conditions the correlation decays exponentially over several particle diameters and the flow and gradient components behave almost identically, indicating that the collective motion is not strongly anisotropic at this level of description. The range of the correlation grows systematically as jamming is approached, in all four cases and for both components (see supplementary material, Fig.~S3). At a fixed distance from jamming, rolling resistance slows the decay: at the largest gaps resolved the correlation retains $\approx0.10$--$0.12$ of its contact value for the rolling-frictional cases against $\approx0.05$ for $\mu_r=0$, so that collective translational motion extends over a longer range once relative rolling is constrained. The variable-$\mu_r$ suspension again follows the $\mu_r=0.3$ case. Both components show sharp steps at the gaps marked by the dashed lines, $h/a_s=2$, $2.8$, $4$, $4.8$ and $5.6$, which are the gaps spanned by one or two intervening particles, of diameter $2a_s$ and $2a_l$ for a small and a large particle. These gaps coincide with the peaks of the pair correlation function $g(h/a_s)$ (supplementary material, Fig.~S4); after the dominant contact peak at $h=h_r$, the most pronounced are those at $2a_s$ and $2a_l$. $g(h/a_s)$ is nearly the same for all four surface conditions, so the steps reflect the packing geometry rather than the rolling constraint. A pair separated by slightly less than one of these gaps cannot accommodate the corresponding aligned chain of touching particles, and is less strongly correlated than an aligned pair at the gap itself, so that the correlation dips just below each line and jumps at it.

The rotational correlations, instead, behave very differently. Panel (c) of Fig.~\ref{f:c_v_w_h} shows that $C_{\tilde{\omega}}(h)$ is appreciable only for gaps $h\lesssim2a_s$ and vanishes beyond $h\approx2a_l$ in every case: rotational coordination is a local property, whereas translational coordination is collective. More importantly, the sign of the correlation at contact depends on the rolling constraint. For sliding friction alone it is negative at contact and positive over gaps up to $h\approx2a_s$. This alternating pattern is expected when neighboring particles roll against one another like gears -- two frictional particles in rolling contact must rotate in opposite directions relative to each other \citep{orsi2026contact}, so that a particle co-rotates with the neighbors of its neighbors. Once rolling resistance is activated the contact value becomes positive: the particles at contact rotate in the same direction. The correlation then peaks again at $h\approx2a_s$, where two particles are bridged by a single small particle, and is absent at larger gaps, so that co-rotation is typically transmitted through at most one intervening particle. The correlation beyond contact is weak, and weaker still for the rolling-frictional cases.

\begin{figure}[!t]
    \centering
    \includegraphics[width=\linewidth]{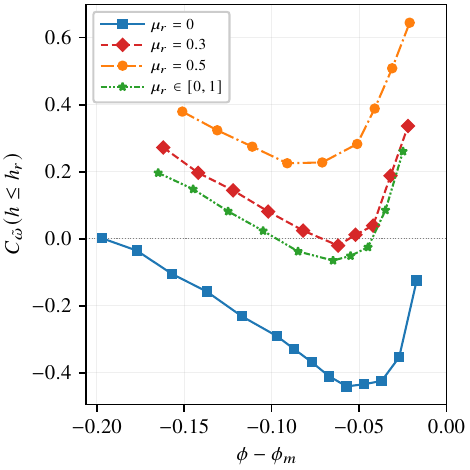}
    \caption{Angular velocity correlation at contact $C_{\tilde{\omega}} (h\leq h_r)$ versus distance from jamming $\phi-\phi_m$ at $\tilde\sigma=100$. Negative values indicate counter-rotation of neighboring particles and positive values co-rotation.}
    \label{f:cw_contact}
\end{figure}

Fig.~\ref{f:cw_contact} follows the contact value $C_{\tilde{\omega}}(h \leq h_r)$ as jamming is approached, at $\tilde\sigma=100$. At $\mu_r=0$ it is strongly negative, reaching $\approx-0.44$, confirming pronounced counter-rotation, in line with the increasingly negative near-contact correlation reported for sliding-frictional suspensions at high stress and density \citep{orsi2026contact}. Rolling friction raises this minimum monotonically: its lowest value is $\approx-0.02$ at $\mu_r=0.3$ and stays above $\approx+0.2$ at $\mu_r=0.5$, while the variable-$\mu_r$ suspension again follows the $\mu_r=0.3$ case, both in the near-contact value and in its dependence on packing fraction. The rolling constraint therefore inverts the local rotational order, from counter-rotating to co-rotating neighbors acting as a rigid unit, a direct kinematic signature of relative rolling being suppressed at contact. Over the full $(\tilde\sigma, \phi)$ plane (supplementary material, Fig.~S5) the same inversion holds wherever the sliding-only suspension counter-rotates at contact, e.g. close to jamming. At the lowest stresses, by contrast, all four surface conditions co-rotate at contact.

All four curves share a common dependence on packing fraction: $C_{\tilde{\omega}}(h \leq h_r)$ passes through a minimum and then rises steeply as $\phi\to\phi_m$, so that close to jamming co-rotation grows regardless of the rolling-friction coefficient, consistent with particles being incorporated into increasingly rigid, collectively-rotating clusters \citep{orsi2026contact,pandare2026contact,santra2025rigid,rahbari2021fluctuations}. The minimum moves away from jamming as the rolling-friction coefficient increases, from $\phi-\phi_m\approx-0.055$ at $\mu_r=0$ to $\approx-0.065$ at $\mu_r=0.3$ and $\approx-0.075$ at $\mu_r=0.5$. The origin of this shift is not clear from the present data and warrants further investigation.

\subsection{Granular temperature}

\begin{figure*}
    \centering
    \includegraphics[width=\linewidth]{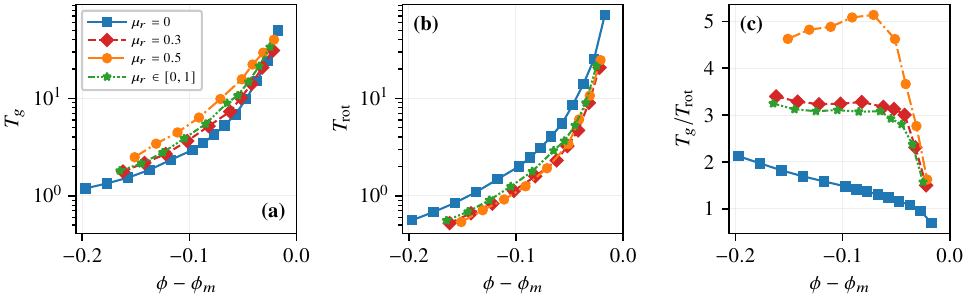}
    \caption{Granular temperatures as a function of the distance from jamming at $\tilde{\sigma}=100$, for the four surface conditions: (a) translational, $T_g$, (b) rotational, $T_{\rm rot}$, and (c) their ratio $T_g/T_{\rm rot}$.}
    \label{f:gtemp}
\end{figure*}

The distributions of the previous section characterize the fluctuations at a single state point. Their second moments provide a more compact measure, which can be followed across the flow diagram and compared between the translational and rotational degrees of freedom. We therefore introduce the translational and rotational granular temperatures
\begin{subequations}
\begin{equation}
    T_g = \frac{1}{2}\left(\langle v_x'^2 \rangle + \langle v_y'^2 \rangle\right) \ ,
\end{equation}
\begin{equation}
    T_{\rm rot} = \langle \omega'^2 \rangle \ ,
\end{equation}
\end{subequations}
Above, $\langle \cdot \rangle$ is both the ensemble and the temporal average, and $v_x'$, $v_y'$ and $\omega'$ are the normalized non-affine velocities introduced above.

Both temperatures are built from the normalized velocities, so each measures the fluctuation relative to the imposed flow. Since both $\omega'$ and $v'$ are dimensionless, the two temperatures are directly comparable. The ratio $T_g/T_{\rm rot}$ then carries no factor of $\dot{\gamma}$ and can be compared across packing fractions, stresses and surface conditions.

Panels (a) and (b) of Fig.~\ref{f:gtemp} show the two granular temperatures as a function of the distance from jamming at $\tilde{\sigma}=100$. Both rise steeply as $\phi\to\phi_m$: measured against the imposed flow, motion becomes increasingly non-affine in the rotational degree of freedom as much as in the translational one. The two temperatures respond to surface conditions in opposite senses. The translational temperature increases with the rolling-friction coefficient, the sliding-only suspension lying lowest over most of the range, whereas the rotational temperature decreases with it, the sliding-only suspension lying above all three rolling-frictional cases at every packing fraction studied while the rolling cases cluster closely together. Rolling resistance therefore suppresses rotational fluctuations while leaving translational ones, if anything, slightly enhanced, consistent with the narrowing of $P(\omega')$ reported above.

The ratio $T_g/T_{\rm rot}$, in panel (c) of Fig.~\ref{f:gtemp}, makes the asymmetry between the two degrees of freedom explicit. For $\mu_r=0$, $T_g/T_{\rm rot}$ decreases monotonically while approaching jamming, so that the rotational temperature rises faster than the translational one and the two become comparable as jamming is approached. The rolling-frictional suspensions behave differently in kind: the ratio stays between $\approx3$ and $\approx5$ over most of the range and drops only in the last few points. For $\mu_r=0.5$ it is not monotonic but passes through a maximum of $\approx5.1$ near $\phi-\phi_m\approx-0.075$, close to where the contact value of the angular velocity correlation has its minimum. At a fixed distance from jamming of $\phi-\phi_m=-0.1$ the sliding-only value is $\approx1.5$ against $\approx5.0$ at $\mu_r=0.5$, a factor of more than three. With a rolling constraint the two temperatures therefore rise at similar rates over most of the range and separate only near $\phi_m$, whereas without rolling resistance the rotational temperature outgrows the translational one throughout, with the ratio falling steadily from the farthest state from jamming studied. In all four cases the ratio is still falling steeply at the closest approach we resolve: the sliding-only suspension has already passed $T_g/T_{\rm rot}=1$, and the rolling-frictional cases would be expected to do so at packings closer to jamming. Whether the ratio tends to a finite value or to zero as $\phi\to\phi_m$ cannot be decided from these data. Rolling friction therefore appears to delay the crossover to rotationally-dominated fluctuation rather than to prevent it.

Taken with the angular velocity distributions, this supports the picture set out in the introduction. Approaching shear jamming, fluctuations grow in both degrees of freedom when measured against the imposed flow, the rotational contribution growing faster \citep{orsi2026contact}; in the sliding-only suspension the two become comparable at the closest approach to jamming we can resolve. Rolling friction, acting directly on the relative rotation at contact, holds the rotational contribution down and delays the point at which it comes to dominate. Conventional diagnostics of shear jamming -- a diverging viscosity, an isostatic contact number -- identify the transition equally well in all four systems, but they are insensitive to this asymmetry between the two degrees of freedom.

\begin{figure*}
    \centering
    \includegraphics[width=\linewidth]{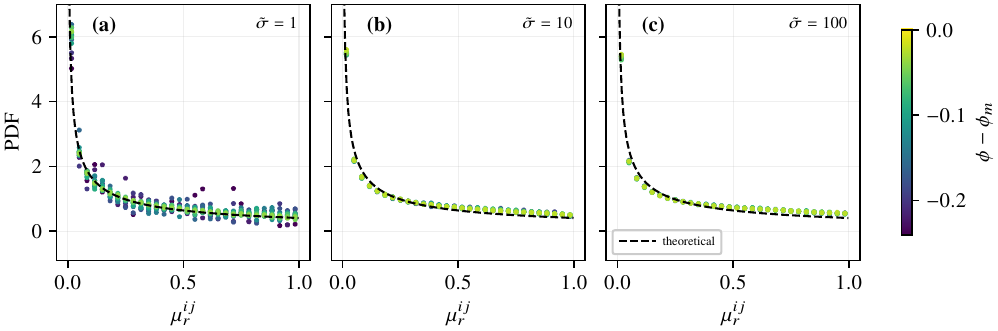}
    \caption{Probability density function of the effective rolling-friction coefficient $\mu_r^{ij}$ at contact between two particles $i$ and $j$, at (a) $\tilde\sigma=1$, (b) $\tilde\sigma=10$ and (c) $\tilde\sigma=100$, and at various distances from jamming indicated. Black dashed lines show the random-pairing prediction $P_{\mathrm{contact}}(m)$. The measured PDF follows the predicted shape but lies somewhat above it at large $\mu_r^{ij}$.}
    \label{f:pdf_mur}
\end{figure*}

\begin{figure}[!t]
    \centering
    \includegraphics[width=\linewidth]{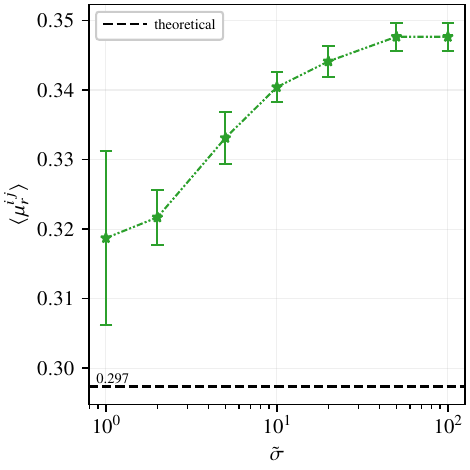}
    \caption{Mean effective rolling-friction coefficient at contact for the variable-$\mu_r$ suspension, as a function of imposed stress; error bars show the spread across packing fraction. The dashed line marks the random-pairing prediction $\langle\mu_r^{ij}\rangle \approx 0.297$, which the measured mean exceeds at every condition, rising with stress and saturating near $0.35$.}
    \label{f:local_mur}
\end{figure}

The granular temperatures and velocity correlations characterize, respectively, the magnitude and spatial organization of the non-affine motion. These kinematic quantities are also connected to dissipation and rheology. From the steady-state energy balance, \citet{trulsson2017effect} showed that the mean-square translational and rotational velocity fluctuations contribute to viscous dissipation, while sliding at frictional contacts provides a separate dissipation mechanism that can dominate close to jamming. In the present system, the opposite variations of $T_g$ and $T_{\rm rot}$ with $\mu_r$ therefore indicate that rolling resistance redistributes non-affine motion, and hence the associated viscous dissipation, between translation and rotation. The granular temperatures alone, however, do not determine how the total dissipation or viscosity changes, because they do not resolve lubrication losses or dissipation through sliding and rolling at contacts. Establishing whether the suppression of rotational fluctuations is compensated by another dissipation channel would require a full decomposition of the stress or power balance. The differences in $T_g/T_{\rm rot}$ between the four surface conditions narrow at the closest distances to jamming that we resolve, but the present data do not determine whether they vanish asymptotically.

\subsection{Effective surface rolling resistance}\label{sec:mur_eff}

We now turn to the suspension with variable rolling friction. Because the rolling friction coefficient varies over each particle surface, the variable-$\mu_r$ suspension is not characterized by the coefficient of an individual particle but by the value actually realized at a contact. We determine this effective coefficient here and use it to interpret the rheology reported above; the model of Sec.~\ref{sec:varmur} predicts $\langle\mu_r^{ij}\rangle\approx0.297$ under random pairing.

Fig.~\ref{f:pdf_mur} shows that the measured distribution reproduces the predicted shape, including the strong weighting towards small $\mu_r^{ij}$, though it lies somewhat above the random-pairing curve at large $\mu_r^{ij}$. The corresponding mean friction coefficient, reported in Fig.~\ref{f:local_mur} versus the applied shear stress, therefore exceeds $0.297$ at all conditions, rising with the imposed stress from $\approx0.32$ at $\tilde\sigma=1$ to $\approx0.35$ for $\tilde\sigma\gtrsim50$, beyond which it saturates and only weakly depends on the packing fraction. The measured value thus lies between the random-pairing prediction and the $0.5$ expected for perfectly correlated pairing, indicating that the contact network does not sample the surfaces uniformly.

The variable-$\mu_r$ suspension is thus characterized by an effective rolling-friction coefficient of $\approx0.35$ at the stresses where thickening and jamming occur. This accounts for the behavior reported above: in its viscosity and jamming point, its flow curve, its granular temperatures, and in the distribution and near-contact correlation of the angular velocity, the variable-$\mu_r$ suspension follows the uniform case $\mu_r=0.3$ rather than $\mu_r=0.5$. This follows from the contact model: because a contact always takes the smaller of the two local coefficients (Sec.~\ref{sec:varmur}), it realizes the smoother of the two surfaces in contact. The equivalence is nonetheless macroscopic: the variable-$\mu_r$ suspension reproduces the bulk response of a uniform $\mu_r\approx0.3$ suspension even though its contacts sample a broad distribution of coefficients rather than a single value. This is a surprising but important result. Whether it generalizes, for example to mixtures of particles with different friction coefficients, remains an open question, as does whether the difference is visible in local observables or whether coefficients at neighboring contacts are correlated.

\section{Conclusion}\label{sec:conclusion}

We have examined how rolling friction affects the rheology and particle kinematics of frictional dense suspensions approaching shear jamming. Stress-controlled two-dimensional discrete-element simulations were performed for a sliding-only reference, $\mu_r=0$, two uniform rolling-friction coefficients, $\mu_r=0.3$ and $0.5$, and a system in which $\mu_r$ varies over each particle surface.

Rolling resistance lowers the high-stress jamming packing fraction from $\phi_m\approx0.80$ at $\mu_r=0$ to approximately $0.76$ and $0.75$ at $\mu_r=0.3$ and $0.5$, respectively. At the fixed packing fraction $\phi=0.76$, this shift changes continuous shear thickening into discontinuous shear thickening and eventually brings the rolling-frictional systems to shear jamming. The separation between the flow curves is much smaller when the systems are compared at the same $\Delta\phi=\phi-\phi_m(\sigma)$. Thus, rolling friction promotes shear thickening at fixed $\phi$ largely by moving the stress-dependent jamming point, rather than by producing a different rheology at a given distance from jamming.

This near-collapse provides the basis for the kinematic comparison. We examined particle motion mainly at $\tilde{\sigma}=100$, where the suspensions are in the high-stress thickened state, and compared systems at matched $\Delta\phi$. Differences between them can then be associated with the rolling constraint rather than with one system simply being closer to jamming than another. The contact network already reflects this constraint: as jamming is approached, the rolling-frictional systems require fewer frictional contacts than the sliding-only system.

The strongest kinematic effect appears in particle rotation. With sliding friction alone, contacting particles counter-rotate strongly. Rolling friction shifts the contact correlation toward co-rotation, with the change becoming more pronounced as $\mu_r$ increases. In every system, the contact correlation turns upward close to jamming, indicating that neighboring particles increasingly rotate together as collective structures develop. Rolling friction therefore changes the local rotational organization; it does not merely reduce the magnitude of rotational motion.

Translational and rotational correlations remain different in spatial extent. Translational velocity correlations persist over several particle diameters, while rotational correlations vanish beyond approximately one particle diameter from contact. At matched $\Delta\phi$, the normalized translational correlations remain stronger at large separations when rolling friction is present, although the data do not establish a different decay law. Rotational correlations, by contrast, are controlled primarily by the local contact constraint.

The fluctuation amplitudes show the same separation between translation and rotation. Both translational and rotational granular temperatures increase on approaching jamming. Without rolling friction, the rotational temperature grows faster and exceeds the translational temperature at the closest state resolved. Rolling friction suppresses rotational fluctuations and leaves the translational temperature larger throughout the accessible range. Even then, $T_g/T_{\mathrm{rot}}$ decreases sharply near jamming, showing that rotation becomes increasingly important in all cases. The present data do not determine whether the rolling-frictional systems eventually become rotationally dominated or whether the ratio approaches a finite value.

The system with surface-varying rolling friction follows the uniform $\mu_r=0.3$ system in the rheological and kinematic observables considered here, despite having a surface-averaged coefficient of $0.5$. Under the chosen contact law, each contact takes the smaller of the two local coefficients. The resulting mean contact value rises with stress and approaches approximately $0.35$ in the thickened regime. The contact distribution is close to, but not identical to, the random-pairing prediction, with a modest excess of large coefficients. These results show that the surface average is not the relevant measure of rolling resistance in this model. The response is instead consistent with the lower coefficients selected at contacts, while the remaining departure from random sampling points to correlations that require contact-resolved analysis.

Bulk rheology therefore does not give a complete account of the effect of rolling friction. Scaling by $\Delta\phi$ captures much of its influence on shear thickening, yet systems with similar flow curves differ markedly in their translational and rotational motion. Reduced models of rough or angular particles should consequently be assessed not only through viscosity and jamming, but also through the contact-scale kinematics that the rolling constraint directly controls.

\section*{Supplementary Material}

See the supplementary material for the state diagrams in the $(\tilde\sigma,\phi)$ plane, the macroscopic friction coefficient, the translational velocity correlations and pair correlation function at all simulated packing fractions, and contour maps of the angular-velocity correlation at contact.

\begin{acknowledgments}
Naveen Agrawal acknowledges the computational resources provided by the National Academic Infrastructure for Supercomputing in Sweden (NAISS), partially funded by the Swedish Research Council through grant agreement no.~2022-06725.
\end{acknowledgments}

\section*{Author Declarations}

\subsection*{Conflict of Interest}

The authors have no conflicts to disclose.

\subsection*{Data Availability}

The data that support the findings of this study are available upon reasonable request.

\bibliography{bibliography}

\end{document}